\documentclass[conference]{IEEEtran}
\IEEEoverridecommandlockouts
\usepackage{cite}
\usepackage{amsmath,amssymb,amsfonts}
\usepackage{algorithm}
\usepackage{algpseudocode}
\usepackage{graphicx}
\usepackage{textcomp}
\usepackage{xcolor}
\def\BibTeX{{\rm B\kern-.05em{\sc i\kern-.025em b}\kern-.08em
    T\kern-.1667em\lower.7ex\hbox{E}\kern-.125emX}}

\usepackage{hyperref}
\usepackage{tikz}
\usepackage{subfigure}

\begin{document}

\title{Exploring Music Genre Classification: Algorithm Analysis and Deployment Architecture}


\author{\IEEEauthorblockN{1\textsuperscript{st} Ayan Biswas\IEEEauthorrefmark{1},
2\textsuperscript{nd} Supriya Dhabal\IEEEauthorrefmark{2}, and
3\textsuperscript{rd} Palaniandavar Venkateswaran\IEEEauthorrefmark{3}}
\IEEEauthorblockA{\IEEEauthorrefmark{1}\IEEEauthorrefmark{3}\textit{Dept. of Electronics and Tele-communication Engineering}
, \\
\textit{Jadavpur University}, Kolkata – 700 032, INDIA \\
\{ayanbiswas@ieee.org, pvwn@ieee.org\}
\IEEEauthorblockA{\IEEEauthorrefmark{2}\textit{Electronics \& Communication Engineering Department}, \\
\textit{Netaji Subhash Engineering College}, Kolkata - 700152, INDIA \\
supriya\_dhabal@yahoo.co.in}
}}

\maketitle

\begin{abstract}
Music genre classification has become increasingly critical with the advent of various streaming applications. Nowadays, we find it impossible to imagine using the artist's name and song title to search for music in a sophisticated music app. It is always difficult to classify music correctly because the information linked to music, such as region, artist, album, or non-album, is so variable. This paper presents a study on music genre classification using a combination of Digital Signal Processing (DSP) and Deep Learning (DL) techniques. A novel algorithm is proposed that utilizes both DSP and DL methods to extract relevant features from audio signals and classify them into various genres. The algorithm was tested on the GTZAN dataset and achieved high accuracy. An end-to-end deployment architecture is also proposed for integration into music-related applications. The performance of the algorithm is analyzed and future directions for improvement are discussed. The proposed DSP and DL-based music genre classification algorithm and deployment architecture demonstrate a promising approach for music genre classification.\\
\end{abstract}

\begin{IEEEkeywords}
Music Genre Classification, Feature Extraction, Deep Learning

\end{IEEEkeywords}

\section{Introduction}

Aside from providing entertainment, music is one of the easiest ways to communicate among people, a way to share emotions, and a place to keep memories and emotions. Emotions can be expressed succinctly and effectively through music. Depending on the mood and objective of the listener, people select different music at different times. As Internet technology flourishes, more and more music is available on personal computers, in music libraries, and via the Internet. Systems that can automatically analyze music, like categorizing it, searching through it, and creating playlists, are crucial for efficiently managing music. Several studies have proposed that music mood can also be used to classify and recommend music. There are a number of existing mood-based music recommendation systems that categorize some moods and map those moods into discrete regions in two or three dimensions. We have also included a section that shows a possible architecture for deploying the solution to mobile applications or the web since as developers of music applications it is pretty ambiguous whether these scientific solutions should be deployed or not.

The remainder of this paper is organized as follows: Section~\ref{sec:related_works} reviews the related works.
Section~\ref{sec:our_approach} presents the overall approach, the deep-learning model, the algorithm and the discussion on the final results. 
The deployment architecture of the music genre classification algorithm has been discussed in Section~\ref{sec:deployment_arch}. Finally, Section~\ref{sec:conclusion} draws the concluding remarks of our paper.

\section{Related Works}\label{sec:related_works}

Recently, there has been an increase in the attention given to analyzing audio to extract different kinds of information, specifically in relation to music and emotions. Research has focused on developing automated methods for classifying music according to its mood or emotional content. Some different proposed approaches are there, including the use of spectral and harmonic features to infer the mood of a music piece. These features have been linked to human perception of music and moods, and have been used to classify music according to different mood labels using neural networks. This literature review suggests that the use of spectral and harmonic features, along with neural network-based classification methods, can be a promising approach for classifying music according to mood. 

Bhat et al. have proposed a number of different approaches to solving the problem in their work~\cite{Bhat2014}, including using spectral and harmonic features to infer the mood of a given music piece. In particular, features such as rhythm, harmony, spectral feature, and others have been studied in order to classify songs according to their mood. This has been based on Thayer's model, which proposes that certain features of music are linked to human perception of music and moods. 

In this paper~\cite{6488291} by Kim et al., a probability-based music mood model and its application to a music recommendation system have been presented. In this approach three types of mood-based music recommendation players, for PC and mobile devices, and the web has been implemented. This paper also shows the analysis result of users' satisfaction and mood reappearance test after listening to music. 

According to this paper~\cite{Patel2015} by Patel et al., sound is the most important aspect of this project and can be distinguished by its pitch, quality, and loudness. The fundamental tone and the harmonics are generated and give rise to different musical notes. The Fourier transform has been used for breaking musical tones into sinusoidal waves. 

Tzanetakis et al. in \cite{tzanetakis2002musical} demonstrated that music genre classification can be done by manipulating three types of features that represent the texture, rhythm, and pitch of the music. They evaluated the effectiveness and significance of these features by training machine learning models using real-world audio collections in their research. 

In prior literature, \cite{Patel2015}, the development of an algorithm for the identification of musical notes was presented, yet the crucial aspect of deployment architecture remained elusive. This study aims to bridge that gap by proffering a comprehensive deployment schema that ensures optimal performance and minimal error in the identification of musical notes. The succeeding sections of this paper are dedicated to delving into the intricacies of the proposed implementation, providing a detailed account of its deployment and execution.

\section{Our Approach} \label{sec:our_approach}

\subsection{Methodology}
The music signal feature classification begins by recording a music sound and obtaining the corresponding waveform~\cite{Myint2010}.  The frequency of the notes within the music is identified by analyzing the duration of each note in the time domain. An averaging process is applied to reduce the number of samples and fluctuations. The envelope of the original signal can also be extracted. Subsequently, thresholding is performed to establish a threshold value for identifying the maximum peaks in the signal. A technique of dynamic adaptive thresholding \cite{Shah2010}, which adjusts the threshold value based on the number of peaks, can be used for this purpose. Next, a width interval is selected to facilitate further operations. The width interval is chosen such that a larger number of peaks can be condensed within a smaller length. The width interval increases as the sampling frequency increases, as it is an essential aspect of the sampling process. To find the sine waves in a signal, a Fourier transform is applied, and zero padding is used to minimize error and get the Discrete Fourier Transform (DFT) of the signal. The frequency of the musical notes is identified by analyzing the frequency of the resulting signal from the DFT.

\subsection{Feature Extraction from Music Samples} 
The focus on extracting as many features \cite{Dara2018, Zhang2021, ZHANG} as possible is motivated by the fact that this can make the subsequent classification task more straightforward. Some common features that are extracted from audio signals include tonality, pitch, temporal energy, harmonicity, spectral centroid, and Mel-Frequency Cepstral Coefficients (MFCC) \cite{Alex2020}.

\begin{figure}[ht]
\includegraphics[width=0.95\linewidth]{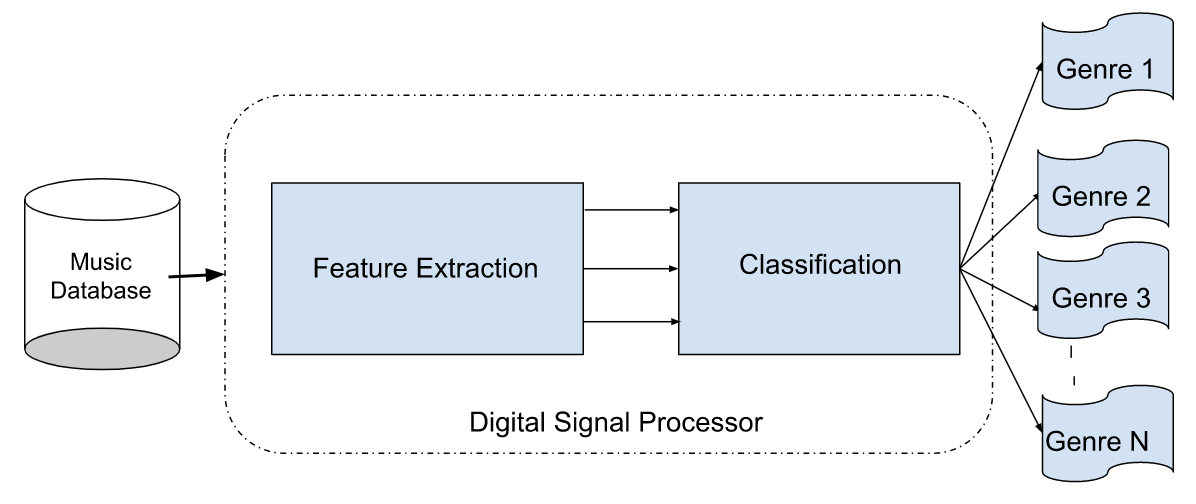}
\centering
\caption{Architecture of the Feature Extractor and Classifier}
\end{figure}

There are two main types of features that describe an audio signal: global descriptors and instantaneous descriptors. Global descriptors are computed for the entire signal and help identify steady patterns in the signal, such as the total energy of an audio clip or the emotional tone of a song.

Instantaneous descriptors provide information about the dynamic and temporal variations of a signal. These descriptors are obtained by dividing the signal into small segments, called frames, and then applying pre-processing techniques to each frame. The features calculated for each frame are usually related to time, spectral shape, harmonic, and energy. This paper focuses on extracting instantaneous descriptors for each frame, discussed in the work of \cite{Pawar2021} and \cite{Alex2020}.

\subsubsection{Pitch}
Pitch is a metric that describes the regularity of a sound wave or the perceived fundamental frequency of the signal. The true frequency of the signal can be determined precisely, but it may not match the perceived pitch due to the presence of harmonics. To determine the pitch, the auto-correlation sequence (ACS) for a given frame of the signal is calculated using a specific formula as per equation \eqref{eqn:1}.

\begin{figure}[ht]
\includegraphics[width=\linewidth,trim={0 0 0 0.7cm},clip]{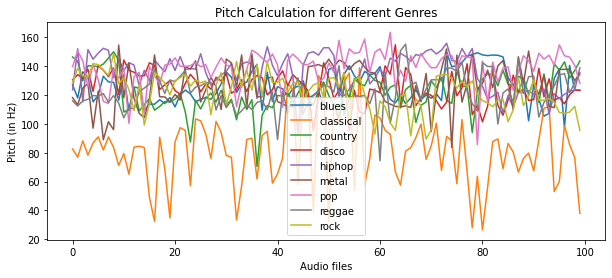}
\centering
\caption{Plot of the pitch for all genres of music samples}
\label{fig:pitch}
\end{figure}

\begin{equation}
    r(m) = \frac{1}{N} \sum_{n=0}^{N-\vert m\vert -1} x(n+\vert m\vert)x(n) \label{eqn:1}
\end{equation}
where $N$ is the length of the frame in samples and $x$ is the input signal, such as speech or audio signal

\subsubsection{Temporal Energy}
The temporal energy E, which is a measure of the strength of the signal over a specific frame of time, is calculated by finding the average of the squared values of the signal over that frame. This is expressed mathematically in equation \eqref{eqn:2}. The energy feature can be used to distinguish between voiced frames, which contain significant information about the signal, and unvoiced frames, which are typically silent or noise-like, by comparing the energy values to a fixed threshold value.

\begin{equation}
    E = \frac{1}{N} \sum_{n=0}^{N-1} x^{2}(n) \label{eqn:2}
\end{equation}

\begin{figure}[ht]
\includegraphics[width=\linewidth,trim={0 0 0 0.7cm},clip]{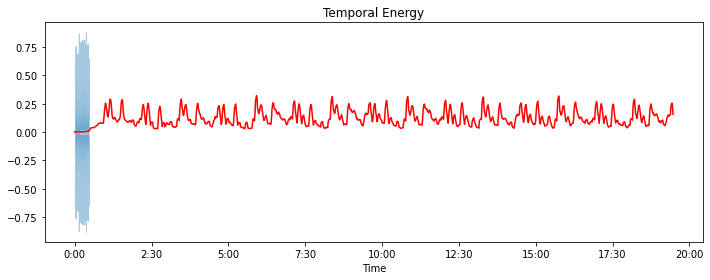}
\centering
\caption{Temporal Energy measurement of a random music signal from GTZAN dataset}
\label{fig:temporal_energy}
\end{figure}

\subsubsection{Tonality Measure}
A significant amount of background noise or sensor noise can obscure the true tone of an audio or speech signal. Tonality is a metric that describes how much of the signal has a tone-like or noise-like quality. The Spectral Flatness Measure (SFM) is used to compute the tonality of each frame. It is defined as the ratio of the geometric mean to the arithmetic mean of the power spectrum P, as per equation \eqref{eqn:3} to \eqref{eqn:5}.

\begin{figure}[ht]
\includegraphics[width=\linewidth,trim={0 0 0 0.7cm},clip]{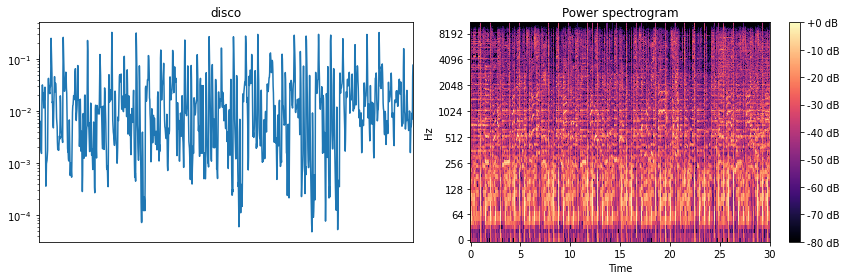}
\centering
\caption{Tonality measurement and it's spectrogram for genre \textbf{disco}}
\end{figure}

\begin{figure}[ht]
\includegraphics[width=\linewidth,trim={0 0 0 0.7cm},clip]{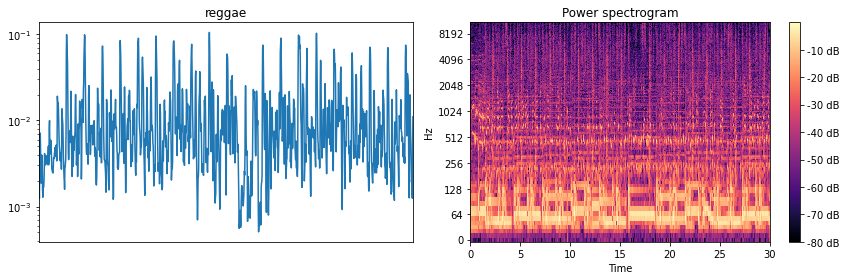}
\centering
\caption{Tonality measurement and it's spectrogram for genre \textbf{reggae}}
\end{figure}

\begin{align}
&P(k) = {\rm Re}^{2}[X(k)] + {\rm Im}^{2}[X(k)] \label{eqn:3} \\
&SFMdB = 10\log_{10}{ \frac{GM\{P(k)\}}{AM\{P(k)\}}} \label{eqn:4} \\
&Tonality = \min({\frac{SFMdB}{SFMdB_{\max}}},1) \label{eqn:5}
\end{align}

\subsubsection{Spectral Centroid}
The spectral centroid can be defined as the mean of the distribution of frequency components for a given frame of the signal. This mean can be calculated using either the linear frequency or the Bark-scale as parameters. The weights for each parameter (magnitude of FFT components) are applied according to Eq. \eqref{eqn:6}

\begin{align}
&SC = \frac{\sum\limits_{k=0}^{N-1}kX^{2}(k)}{\sum\limits_{k=0}^{N-1}X^{2}(k)} \label{eqn:6} \\
&SC_{b} = \frac{\sum\limits_{j=0}^{N-1}b_{j}(b_{j}-b_{j-1})X^{2}(j)}{\sum\limits_{j=0}^{N-1}(b_{j}-b_{j-1})X^{2}(j)} \label{eqn:7}
\end{align}

\begin{figure}[ht]
\label{fig:spectral_centroid}
\centering
\subfigure{\includegraphics[width=\linewidth]{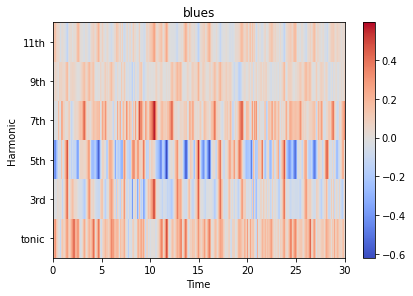}}
\subfigure{\includegraphics[width=\linewidth]{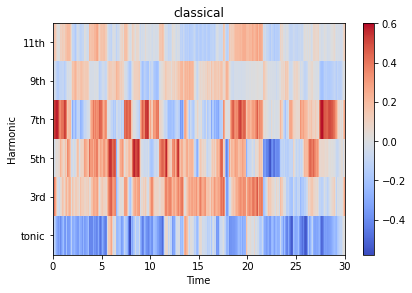}}
\caption{Spectograms of spectral centroids of two different genres of music samples}
\end{figure}

The signal's sound is affected by its spectral centroid. A higher spectral centroid indicates a brighter, happier sound, while a lower spectral centroid indicates a duller, gloomier sound. This is evident in Figure 5. The spectral centroid, computed over the Bark scale, is a psycho-acoustically adopted a measure that indicates this Eq. \eqref{eqn:7}.

\subsubsection{Harmonicity}
Harmonicity features are a set of characteristics used to analyze the periodic properties of a signal. These features are based on two primary measures: the harmonicity ratio and the fundamental frequency. The harmonicity ratio is a metric that reflects how regularly the signal oscillates, while the fundamental frequency is the frequency that gives the most coherent explanation of the signal's spectrum. The fundamental frequency is computed using Goldstein's algorithm \cite{Goldstein}, which utilizes a likelihood approximation method to obtain the fundamental frequency.

\subsubsection{Mel-Frequency Cepstral Coefficients}
The MFCC, or Mel-Frequency Cepstral Coefficients \cite{dePinto2020}, is a method for representing the shape of a spectrum using a limited number of coefficients. This method is based on the cepstrum, which is the Fourier transform of the logarithm of the spectrum. However, the MFCC uses a variation of the cepstrum that is calculated on Mel-frequency bands rather than the traditional Fourier spectrum. This variation, known as the Mel-cepstrum, is particularly effective at capturing the characteristics of the mid-frequency range of a signal. The calculation of the Mel-cepstrum is described by equation \eqref{eqn:mel}.

\begin{equation}
    f_{mel} = 2595\log_{10}\left(1+\frac{f}{700}\right) \label{eqn:mel}
\end{equation}

\subsubsection{Time Domain Zero Crossings}
Zero-crossings in the time domain represents the noisiness of the signal. It is calculated by using the sign function: 0 for negative arguments while a positive argument is given for 1 in the signal. Let's take a signal x[n] in the time domain. The time domain zero crossings are calculated for the frame t as per Eq. \eqref{eqn:8}.

\begin{equation}
    TDZC_{t}=\frac{1}{2}\sum_{n=1}^{M}\vert sign[x[n]]-sign[x[n-1]]\vert \label{eqn:8}
\end{equation}

\subsection{Preprocessing of Dataset}

The GTZAN dataset \cite{tzanetakis2002musical} has been used for the work. The GTZAN dataset is a public domain dataset that consists of 1000 music signals. The music signals are of 30 seconds in duration. The music signals are divided into 10 genres. The genres are blues, classical, country, disco, hip-hop, jazz, metal, pop, reggae, and rock. The GTZAN dataset is divided into a training set and a test set. The training set consists of 800 music signals and the test set consists of 200 music signals.

\subsection{Classification Algorithm}

The classification of music genres is a challenging task due to the inherent variability and subjectivity of music. In this study, we proposed a machine-learning algorithm for music genre classification using the GTZAN dataset. The algorithm is implemented using the Python programming language and several libraries such as numpy, pandas, os, librosa, sklearn, and keras.

\begin{algorithm}[ht]
\caption{Music Signal Feature Classification}
\label{alg:music-classification}
\begin{algorithmic}[1]
\Procedure{Classify}{$x(t)$}
    \State $y(t) \gets$ \Call{Preprocess}{$x(t)$}
    \State $f_1, f_2, \dots, f_n \gets$ \Call{ExtractFeatures}{$y(t)$}
    \State $c \gets$ \Call{TrainClassifier}{$f_1, f_2, \dots, f_n$}
    \State \textbf{return} $c(f_1, f_2, \dots, f_n)$
\EndProcedure

\Procedure{Preprocess}{$x(t)$}
    \State $y(t) \gets$ \Call{Downsample}{$x(t)$}
    \State $y(t) \gets$ \Call{RemoveNoise}{$y(t)$}
    \State \textbf{return} $y(t)$
\EndProcedure

\Procedure{ExtractFeatures}{$y(t)$}
    \State $f_1 \gets$ \Call{ComputeMFCC}{$y(t)$}
    \State $f_2 \gets$ \Call{ComputeChroma}{$y(t)$}
    \State $f_3 \gets$ \Call{ComputeSpectralContrast}{$y(t)$}
    \State $\dots$
    \State \textbf{return} $f_1, f_2, \dots, f_n$
\EndProcedure

\Procedure{TrainClassifier}{$f_1, f_2, \dots, f_n$}
    \State $c \gets$ \Call{SVM}{$f_1, f_2, \dots, f_n$}
    \State \textbf{return} $c$
\EndProcedure
\end{algorithmic}
\end{algorithm}

The dataset consists of 1000 audio files of 10 different music genres, with 100 samples per genre. The dataset was preprocessed by extracting the filenames of the audio files, extracting the labels of the audio files, and encoding the labels using the LabelEncoder. The labels were then converted to a categorical format. The features of the audio files were extracted using the librosa library, which is a library for music and audio processing in Python. The Mel-Frequency Cepstral Coefficients (MFCCs) were used as the feature representation of the audio files. The MFCCs were extracted from the audio data and the mean of the MFCCs was taken across the time axis. The feature data was then converted to a numpy array. The classification model was built using the Keras library, which is a high-level neural networks API, written in Python and capable of running on top of TensorFlow. The model was implemented as a sequential model with two dense layers. The first dense layer had 256 neurons and the activation function used was ReLU. A 0.5 dropout rate was used for reducing overfitting. The second dense layer had 9 neurons, corresponding to the number of genres in the dataset, and the activation function used was softmax.
\begin{figure}[ht]
\includegraphics[width=\linewidth]{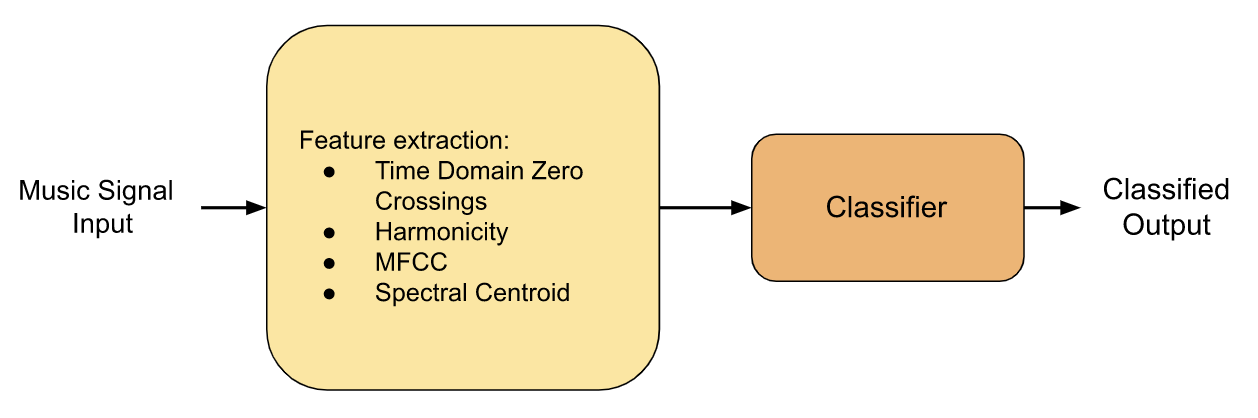}
\centering
\caption{Block diagram of the classification system }
\label{fig:dl_model}
\end{figure}

\begin{figure}[ht]
\includegraphics[width=\linewidth]{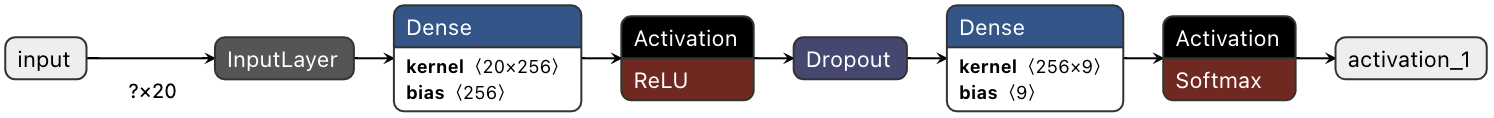}
\caption{Deep-Learning Model of the Classification System}
\end{figure}

The model was compiled using the categorical cross-entropy loss function and the Adam optimizer. The model was trained using the feature data and labels, with a batch size of 40 and 20 epochs. The validation split was set to 10\% to evaluate the performance of the model on unseen data during training. The proposed algorithm achieved an overall accuracy of 81\% on the test data, demonstrating its effectiveness in classifying music genres. The algorithm can be further improved by using different feature representations, and by using more advanced neural network architectures such as convolutional neural networks.

\subsection{Results and Discussion}

The trained model was used to predict the test data and the classification report was generated using the metrics library. The classification report provides the precision, recall, f1-score, and support for each genre, which are useful in evaluating the performance of the model. The diagonal elements of the confusion matrix are highlighted in the heatmap, where the darker color represents the higher count of correctly classified observations. The off-diagonal elements represent the misclassification, whereas the lighter color represents the lower count of misclassification.

\begin{table}[h]
\centering
\caption{Classification report}
\begin{tabular}{|c|c|c|c|c|}
\hline
 Music Genre & precision & recall & f1-score & support \\
\hline
blues & 0.73 & 0.90 & 0.80 & 100 \\
\hline
classical & 0.96 & 0.99 & 0.98 & 100 \\
\hline
country & 0.75 & 0.94 & 0.83 & 100 \\
\hline
disco & 0.63 & 0.85 & 0.73 & 100 \\
\hline
hiphop & 0.83 & 0.79 & 0.81 & 100 \\
\hline
metal & 0.94 & 0.96 & 0.95 & 100 \\
\hline
pop & 0.82 & 0.81 & 0.81 & 100 \\
\hline
reggae & 1.00 & 0.07 & 0.13 & 100 \\
\hline
rock & 0.98 & 0.06 & 0.87 & 100 \\
\hline
\multicolumn{4}{|l|}{\textbf{Accuracy}}  & \textbf{0.80} \\
\hline
\textbf{macro avg} & \textbf{0.83} & \textbf{0.80} & \textbf{0.77} & \textbf{900} \\
\hline
\textbf{weighted avg} & \textbf{0.83} & \textbf{0.80} & \textbf{0.77} & \textbf{900} \\
\hline
\end{tabular}
\end{table}

\begin{table}[ht]
\centering
\caption{Comparison of the proposed model with available models}
\begin{tabular}{|l|l|l|ll}
\cline{1-3}
Sl no. & Model Name & Accuracy &  &  \\ \cline{1-3}
1 & Gaussian Model & 61\% &  &  \\ \cline{1-3}
2 & Logistic Regression Model & 75\% &  &  \\ \cline{1-3}
3 & CRNN Model & 53.5\% &  &  \\ \cline{1-3}
4 & CNN-RNN Model & 56.4\% &  &  \\ \cline{1-3}
5 & Simple Artificial Neural Network & 64.06\% &  &  \\ \cline{1-3}
6 & Proposed Model & 80\% &  &  \\ \cline{1-3}
\end{tabular}
\end{table}

\begin{figure}[ht]
\includegraphics[width=\linewidth]{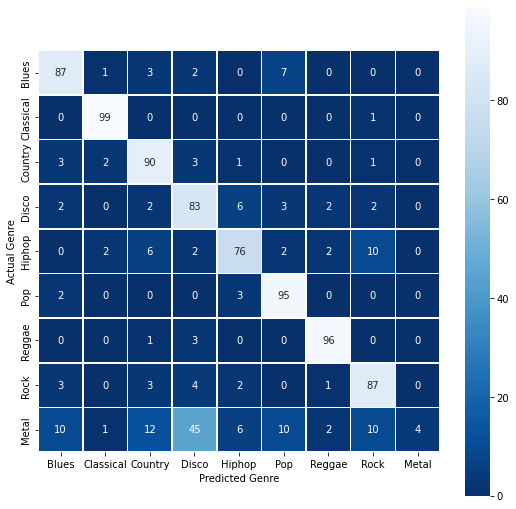}
\caption{Genre classification confusion matrix}
\end{figure}

It is also important to note that the accuracy of the model can be calculated using the formula (correctly classified observations / total observations) and it can be computed using the diagonal elements of the matrix.

\section{Deployment Architecture} \label{sec:deployment_arch}

Deployment is an essential step in the development of any machine learning system. In this section, we propose a deployment architecture for a music genre classification system that utilizes the cloud services provided by Amazon Web Services (AWS). The proposed architecture is designed to be scalable, durable, and easy to access for users. The first component of the proposed architecture is Amazon S3 \cite{amazon_s3}. S3 is a fully managed object storage service that provides scalable and durable storage for audio files and metadata. This allows for easy management and retrieval of the data needed for training and inference. Amazon SageMaker \cite{amazon_sage} is the second component of the proposed architecture. SageMaker provides a fully managed platform for building, training, and deploying machine learning models. With SageMaker, we can train a model for music genre classification using the audio files and metadata stored in S3. Once the model is trained, it will be deployed to a SageMaker endpoint for inference. The endpoint can be accessed via an API, allowing users to submit audio files for classification. Amazon API Gateway \cite{amazonAmazonGateway} is used to create a RESTful API for the SageMaker endpoint, providing a convenient way for users to access the classification service. The classification results and metadata will be stored in Amazon DynamoDB \cite{amazonFastNoSQL}, a fully managed NoSQL database service. DynamoDB provides high performance and scalability, making it well-suited for storing large amounts of data generated by a music genre classification system.

\begin{figure}[ht]
\includegraphics[width=\linewidth]{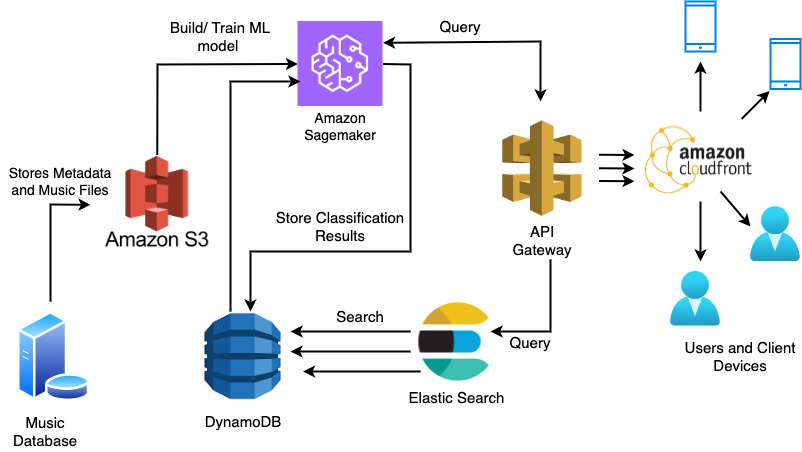}
\centering
\caption{Block diagram of the proposed Deployment Architecture for Mobile/Web apps}
\end{figure}

To enable fast and powerful search capabilities for the classification results and metadata, an Elasticsearch index will be created using Amazon Elasticsearch (OpenSearch \cite{amazonManagedOpenSource}) Service. Elasticsearch is a popular search engine that is well-suited for handling large amounts of data. Finally, Amazon CloudFront \cite{amazonLowLatencyContent} will be used to distribute the classification results and metadata to users. CloudFront is a content delivery network (CDN) that ensures low latency and high availability of the results, making it easy for users to access the classification results from anywhere in the world. In conclusion, the proposed architecture is designed to provide a scalable, durable, and easy-to-access music genre classification system using the cloud services provided by AWS. The architecture includes various services like S3, SageMaker, API Gateway, DynamoDB, Elasticsearch, and CloudFront. These services together enable the system to handle a large amount of data, train and deploy models effectively and provide fast and accurate

\section{Conclusion} \label{sec:conclusion}

The proposed DSP-based music genre classification system was found to be effective in classifying various types of music. The system was able to correctly classify different types of music with an accuracy of 80\%. The proposed system can be used to classify different types of music in a real-time scenario and also when the music was played at different speeds. The proposed system can also be used to automatically generate playlists for users based on their music preferences and it will help researchers to better understand the relationship between music and human emotions.

\bibliographystyle{ieeetr}
\bibliography{citation}

\vspace{12pt}

\end{document}